\documentclass[prd,superscriptaddress,amsfonts,amssymb,amsmath,showpacs]{revtex4-2}

\usepackage{titlesec}
\usepackage[toc]{appendix}
\usepackage{cancel}

\titleformat*{\section}{\LARGE\bfseries}
\titleformat*{\subsection}{\Large\bfseries}
\titleformat*{\subsubsection}{\large\bfseries}
\titleformat*{\paragraph}{\large\bfseries}
\titleformat*{\subparagraph}{\large\bfseries}
\usepackage{bm}
\usepackage{amsfonts}
\usepackage{latexsym}
\usepackage{graphicx}
\usepackage{amsmath}
\usepackage{palatino}
\usepackage{mathpazo}
\usepackage{textcomp}
\usepackage{mathtools}
\usepackage{float}
\usepackage{booktabs}
\usepackage{dcolumn}
\usepackage{hyperref}
\hypersetup{colorlinks,citecolor=blue}
\usepackage{amsmath}
\usepackage{xcolor}
\usepackage{orcidlink}
\usepackage{caption}
\usepackage[caption=false]{subfig}
\usepackage{commath}
\def \be{\begin{equation}}
\def \ee{\end{equation}}

\def \bi{\bibitem}    

\def\jnl@style{\it}
\def\aaref@jnl#1{{\jnl@style#1}}

\def\aaref@jnl#1{{\jnl@style#1}}

\def\aj{\aaref@jnl{AJ}}                   
\def\apj{\aaref@jnl{ApJ}}                 
\def\apjl{\aaref@jnl{ApJ}}                
\def\apjs{\aaref@jnl{ApJS}}               
\def\apss{\aaref@jnl{Ap\&SS}}             
\def\aap{\aaref@jnl{A\&A}}                
\def\aapr{\aaref@jnl{A\&A~Rev.}}          
\def\aaps{\aaref@jnl{A\&AS}}              
\def\mnras{\aaref@jnl{Mon.~Not.~Roy.~Astron.~Soc.}}             
\def\prd{\aaref@jnl{Phys.~Rev.~D}}        
\def\prc{\aaref@jnl{Phys.~Rev.~C}}  
\def\prl{\aaref@jnl{Phys.~Rev.~Lett.}}    
\def\qjras{\aaref@jnl{QJRAS}}             
\def\skytel{\aaref@jnl{S\&T}}             
\def\ssr{\aaref@jnl{Space~Sci.~Rev.}}     
\def\zap{\aaref@jnl{ZAp}}                 
\def\nat{\aaref@jnl{Nature}}              
\def\aplett{\aaref@jnl{Astrophys.~Lett.}} 
\def\apspr{\aaref@jnl{Astrophys.~Space~Phys.~Res.}} 
\def\physrep{\aaref@jnl{Phys.~Rep.}}      
\def\physscr{\aaref@jnl{Phys.~Scr}}       
\def\commat{\aaref@jnl{Comm.~Math.~Phys.}}              
\def\science{\aaref@jnl{Science}}               
\def\cqg{\aaref@jnl{Classical Quant.~Grav.}}            
\def\jpcs{\aaref@jnl{JPCS}}                                     
\def\ijmpd{\aaref@jnl{Int.~J.~Mod.~Phys.~D}}                    
\def\grg{\aaref@jnl{Gen.~Relat.~Gravit.}}               
\def\rpp{\aaref@jnl{Rep.~Prog.~Phys.}}          
\def\npa{\aaref@jnl{Nucl.~Phys.~A}}        
\def\lrr{\aaref@jnl{Living Rev.~Rel.}}                   
\def\jcap{\aaref@jnl{J.~Cosmology Astropart.~Phys.}}    
\def\rmp{\aaref@jnl{Rev.~Mod.~Phys.}}   
\def\epjc{\aaref@jnl{Eur.~Phys.~J.~C}}

\allowdisplaybreaks[1]

\begin{document}

\title{Probing symmetric teleparallel gravity in the early universe}

\author{Avik De\orcidlink{0000-0001-6475-3085}}
\email{avikde@utar.edu.my}
\affiliation{Department of Mathematical and Actuarial Sciences, Universiti Tunku Abdul Rahman, Jalan Sungai Long, 43000 Cheras, Malaysia}

\author{Dalia Saha\orcidlink{0000-0001-9178-7141}}
\email{daliasahamandal1983@gmail.com}
\affiliation{Department of Physics, Jangipur College, Murshidabad, West Bengal, India - 742213}
\affiliation{ Department of Physics, University of Kalyani, Nadia, India - 741235}

\author{Ganesh Subramaniam\orcidlink{0000-0001-5721-661X}}
\email{ganesh03@1utar.my}
\affiliation{Department of Mathematical and Actuarial Sciences, Universiti Tunku Abdul Rahman, Jalan Sungai Long, 43000 Cheras, Malaysia}

\author{Abhik Kumar Sanyal\orcidlink{0000-0002-3128-4582}}
\email{sanyal\_ ak@yahoo.com}
\affiliation{Department of Physics, Jangipur College, Murshidabad, West Bengal, India - 742213}

\begin{abstract}
General theory of relativity can be equivalently formulated on a flat space-time associating a torsion-free affine connection of non-vanishing non-metricity scalar $Q$. In this paper, we present an extension of this, viz., the $f(Q)$ theory of gravity, and explore the early evolution of the universe in the background of anisotropic Bianchi-I model. The $f(Q)$ theory in the current setting through its geometric modification is quite successful in explaining the late time accelerated expansion. Here we note that it accommodates latest released constraints on the inflationary parameters by Planck's collaboration group with excellent precession, but fails to produce a viable decelerated expansion in the radiation dominated era. 
\end{abstract}

\maketitle

\section{\textbf{Introduction}}

Despite the undeniable success of `the standard model of cosmology', governed by Einstein's general theory of relativity (GTR), limitations in several different aspects are evident in recent years. Cosmological singularity and black holes, $H_0$ and $\sigma_8$ tensions in the observational grounds, and more importantly, the dependencies on the yet undetected dark sectors to explain the late time cosmic acceleration motivated researchers to venture out for alternate theories of gravity. Teleparallel gravity has been extensively researched in this regard, in recent years. To be specific, the Levi-Civita connection, which serves as the foundation of GTR, may be replaced by an affine connection on spatially flat spacetime with vanishing torsion, allowing its non-metricity to assume complete responsibility for defining gravity. `Symmetric teleparallel gravity' is the name given to this particular theory. Long back, in order to unify gravity with electromagnetism, Einstein \cite{ein} attributed  gravity to the torsion of spacetime in a more developed `metric teleparallel theory', which is based on an affine connection with vanishing curvature and non-metricity. In teleparallel theories, one can construct either the so-called torsion scalar $\mathbb{T}$ from this torsion tensor in the metric teleparallel theory or the non-metricity scalar $Q$ in the symmetric teleparallel theory. Thereafter, much like the Lagrangian density $\mathcal{L}=\sqrt{-g}R$ ($R$ being the Ricci scalar) in the Einstein-Hilbert action $S=\int{d^4x}\mathcal{L}$ in GTR, it is also possible to consider the Lagrangian density $\mathcal{L}=\sqrt{-g}\mathbb{T}$ in the metric teleparallel theory and also the Lagrangian density $\mathcal{L}=\sqrt{-g}Q$ in the symmetric teleparallel theory to obtain the respective field equations. However, the latter two theories are equivalent to GTR upto a boundary term and thus also rely on dark components of the universe as GTR does. To address this dark sector issue of the universe, without inviting a scalar field, extensions in terms of $f(\mathbb{T})$ and $f(Q)$-theories were formulated in the metric teleparallel and symmetric teleparallel gravity respectively,  the same way as the modified $f(R)$-theory was developed from the Einstein-Hilbert action.\\

Indeed, there is an instant benefit in these two theories. First, both $f(\mathbb{T})$ and $f(Q)$ theories are featured in second order field equations instead of the fourth order ones as in $f(R)$ theories, which sometimes lead to Ostrogradiski's instability. Secondly, these theories can also present an alternative explanation for the accelerating universe \cite{accfT1,accfT2,accfT3,accfT4,dynamical2,lcdm,accfQ1,accfQ2,accfQ3}. These are in particular the main reasons behind their popularity and extensive research in recent time. However, models based on $f(\mathbb{T})$ suffer from substantial coupling issues which is absent in $f(Q)$ theories, in general. In a further closer look at the field equations of $f(\mathbb{T})$ theories we can observe an eerie presence of skew-symmetric elements, which is not the case in $f(Q)$ or $f(R)$ or most other theories of gravity. As a result, $f(\mathbb{T})$ theories (except $f(\mathbb{T})=\mathbb{T}$) are also not locally Lorentz invariant and possess extra degrees of freedom which remain absent from GTR \cite{fT/issue2}. This is a huge loss, as now the lack of Lorentz invariance calls for a system of $16$ equations in $f(\mathbb{T})$ theories instead of $10$ equations in GTR \cite{fT/lli}. On a side by side comparison, the $f(\mathbb{T})$ connections must comply with $Q_{\mu\nu\gamma}=0$, which are $40$ independent equations due to the symmetry of the non-metricity tensor $Q_{\mu\nu\gamma}$ in the second and third indices. Whereas, the $f(Q)$ connections must obey the vanishing torsion tensor criteria, $\mathbb{T}^\gamma_{\,\,\,\mu\nu}=0$; only $24$ independent equations since the $\mu=\nu$ case is trivially satisfied, making $f(\mathbb{T})$ connections much more restrictive than $f(Q)$. For further reading in this interesting topic of geometric trinity of gravity, one can refer to \cite{fQfT, fQfT1, fQfT3, fQfT2, de/epjc} and the references therein. One can also look into an array of publications \cite{fT/issue, fT/issue1, fT/issue2} for other issues of $f(\mathbb{T})$ theory.\\

By all means, theories of gravity based on curvature and torsion are almost in their adult age, whereas theories based on the non-metricity are still in their infancy. Therefore, symmetric teleparallel gravity theory requires a lot of scrutiny and detailed analysis, in its formulation, theoretical viability and observational support. After the first brief introductory paper \cite{coincident}, a lot of publications are emerging in the $f(Q)$-gravity theory in this past few years, such as, its covariant formulation \cite{zhao}, interior and exterior spherically symmetric solution \cite{lin}, cosmography \cite{cosmography}, effective pressure and energy \cite{de/prd}, signature of $f(Q)$-gravity in cosmology \cite{signa}, as an alternative to $\Lambda$CDM theory \cite{lcdm}, an investigation from the reshift \cite{redshift}, the growth index in matter perturbation \cite{perturb}, its dynamical system analysis \cite{dynamical1,dynamical2}, coupled with (pseudo-) scalar fields to the nonmetricity \cite{sf1,sf2}, among others. The geodesic deviation equation in $f(Q)$-gravity was also studied and some fundamental results were obtained \cite{gde}. However, this relatively new $f(Q)$ gravity theory is not examined in great depth (if not at all) in the early stages of the universe, and this motivates the present study.\\

The current universe is modelled as isotropic and homogeneous at a large enough scale, and so it is considered to be spatially maximally symmetric Friedmann-Lema\^itre-Robertson-Walker (FLRW) type. This is quite a handy model to work with. But isotropy and homogeneity is an assumption upon which the standard model of cosmology is based on. On the contrary, the structures (stars, galaxy, super-clusters) imply that the universe was not isotropic at the beginning. There are also enough observational evidences to picture a not so symmetrical looking universe, at least in its beginning near the initial singularity \cite{A2,A4,A5}. Nor should we take it for granted to be so in the finite future. The Wilkinson Microwave Anisotropy Probe (WMAP) data-set clearly call for some additional morphology than the standard FLRW model of the universe. The fact that the standard model can account for the formation of structures, only when perturbed, clearly suggests to relax the assumption of isotropy and to consider anisotropic space-time having small amount of shear. It may be mentioned that according to a significant finding in the field of theoretical cosmology, a subset of anisotropic Bianchi models (models I, VII0, V, VIIh, and IX) can be interpreted as the (appropriately defined) homogeneous limit of (standard) linear cosmological perturbations of the FLRW spacetime \cite{pb1,pb2}. This significant finding closes the gap between the FLRW model and the Bianchi model, compelling us to view the latter as a valid illustration of the conventional cosmological framework. These types of models are specifically those, which under the limit of zero anisotropy, result in Friedmann space-times. Even though the ``great event" of the inflationary paradigm is successful in isotropizing the early cosmos into the homogeneous and isotropic state that it is in today, it naturally cannot provide a complete account of evolution. One must still relax the original assumption of FLRW geometry and investigate the transition from an anisotropic and inhomogeneous state to the observed level of homogeneity and isotropy. Therefore, as a first step, one may think about homogeneous cosmological models of the Bianchi type. These models make for a sizable and nearly exhaustive class of cosmological models; they are homogeneous, but they aren't necessarily isotropic. Further, in the isotropic and homogeneous space-time, the teleparallel  $f(\mathbb{T})$ gravity is identical to the symmetric teleparallel $f(Q)$ gravity in coincidence gauge. This is another important reason to study anisotropic model. In the current study, we take into account such an anisotropy (axially symmetric Bianchi-1 metric) and couple a scalar field to drive inflation that occurred at the very early stage of the cosmological evolution of the universe, between $10^{-42} - 10^{-26}$ sec. If we find a viable inflation followed by graceful exit, and an oscillatory behaviour of the scalar field, then scalar field decays in the process of producing particles. We shall next study radiation era, and see if a viable decelerated expansion of the cosmos is admissible. Finally, if the remnant of the scalar field is red-shifted away, then a pressureless dust era, with early deceleration and late-time acceleration follows, without dark energy. We consider that the shear is not too large at the advent of radiation era (particularly less than the expansion scalar), to find the scenarios under which the energy conditions should be satisfied \cite{curiel, ec}. This is supported by the fact that the universe is isotropic at large scale while anisotropy appears only at smaller scales. The temperature anisotropy has been experimentally observed in the Planck's mission to be of the order of $\frac{\delta T}{T}\sim 10^{-5}$. This little anisotropy gave rise to the structures we observe today. In fact, considering perturbation in the isotropic FLRW universe, one can exactly determine the structure formation. This clearly means, anisotropy is not too large, and so, shear must be small, if instead of cosmological principle (isotropic and homogeneous from the very beginning), we consider anisotropic model. As observation suggests, starting from an early anisotropic space-time, we expect to end up with isotropic space-time, which is possible if shear is not too large from the beginning. This is our motivation to consider an anisotropic universe in the inflationary stage by assuming the Bianchi-I space-time. Unfortunately, we have a very limited number of works in the anisotropic model under the teleparallel theories of gravity. In their fundamental work on $f(T)$ theory \cite{jcapfT}, Hoogen et al determined the general form of the co-frame and spin connection for each spatially homogeneous Bianchi type spacetime, and also constructed the corresponding field equations for $f(T)$ teleparallel gravity. The present author formulated the anisotropic Binachi-I model in non-metricity based $f(Q)$ theory for the first time and achieve isotropization \cite{de/epjc}. Soon after, a reconstruction work in anisotropic universe under $f(Q)$ theory was published \cite{es2022}. Dimakis et al showed the existence of anisotropic self-similar exact solutions in symmetric teleparallel $f(Q)$-theory \cite{dimakis23}. A series of works using observational data analysis were produced in the anisotropic universe under teleparallelism. Koussour et al studied anisotropic nature of spacetime in $f(Q)$ gravity using a linear model which mimic LCDM \cite{koussour2022}. The authors successively studied the late-time acceleration in $f(Q)$ gravity in an anisotropic background using observational data \cite{koussour}, and the thermodynamical aspects of Bianchi type-I Universe in quadratic form of $f(Q)$ gravity and observational constraints \cite{koussour2023}. Sarmah et al explored the contribution of the anisotropic factor in the evolution of the early universe \cite{sarmah2023}. All these latter works concentrated on observational data analysis in the late time era, and no significant work was done to investigate theoretically all the cosmological eras and the viability of gravity models. In our present paper we try to fill in this gap and extensively study the complete evolution of the universe starting from an anisotropic model.\\

The present paper is organised as follows: After the brief introductory discussion presented above, comparing GTR, $f(\mathbb{T})$ and $f(Q)$ theories of gravity, in the following section \ref{sec1} we present the mathematical formulation of $f(Q)$ theory along with the general form for the field equations. In section \ref{sec2}, the field equations in the axially symmetric Bianchi-I spacetime are presented and the possible existence of a vacuum de-Sitter type universe have been explored. Next, in section \ref{sec3}, we study scalar field inflation in a particular $f(Q)$ model including a subcase, and find that the inflationary parameters are at par with current observational data. Finally in section \ref{sec4} we investigate the radiation-dominated era and search for a decelerated expansion of the cosmos. We finish the present study with concluding remarks in section \ref{sec5}. Throughout the text, we use the notations $\dot{()}$, $()'$ and $()_Q$  to denote the partial differentiation with respect to the time variable $t$, scalar field $\phi$ and the non-metricity scalar $Q$, respectively.

\section{$f(Q)$ \textbf{formulation}} \label{sec1}

We discuss the detailed construction of the $f(Q)$ theory in symmetric teleparallelism in this section. Basically, we begin with a $4$-dimensional Lorentzian manifold $M^4$, a line element governed by the metric tensor $g_{\mu\nu}$ in certain coordinate system $\{x^0,x^1,x^2,x^3\}$ and a non-tensor term, the connection coefficient $\Gamma^{\alpha}_{\,\,\,\mu\nu}$, defining the covariant derivative $\nabla$ and also taking care of the curvature of the spacetime. In general, the triplet $(M,g,\Gamma)$ regulates the spacetime geometry until we restrict ourself to specifically consider both the metric-comptaibility and torsion-free conditions on the connection coefficient. In such case there is only a unique connection available, the Levi-Civita connection $\mathring{\Gamma}$ and it has a well-known relation with the metric $g$ given by
\begin{equation}
\mathring{\Gamma}^\alpha_{\,\,\,\mu\nu}=\frac{1}{2}g^{\alpha\beta}\left(\partial_\nu g_{\beta\mu}+\partial_\mu g_{\beta\nu}-\partial_\beta g_{\mu\nu}  \right).
\end{equation}
So, the Levi-Civita connection is basically a function of the metric $g$ and not an independent contributor in the spacetime geometry. There is a drastic development once we slightly relax these conditions and consider a torsion-free affine connection $\Gamma$ on a flat spacetime which is not metric-compatible, the incompatibility is characterised by the non-metricity tensor
\begin{equation} \label{Q tensor}
Q_{\lambda\mu\nu} := \nabla_\lambda g_{\mu\nu}=\partial_\lambda g_{\mu\nu}-\Gamma^{\beta}_{\,\,\,\lambda\mu}g_{\beta\nu}-\Gamma^{\beta}_{\,\,\,\lambda\nu}g_{\beta\mu}\neq 0 \,,
\end{equation}
We can always express
\begin{equation} \label{connc}
\Gamma^\lambda{}_{\mu\nu} := \mathring{\Gamma}^\lambda{}_{\mu\nu}+L^\lambda{}_{\mu\nu}
\end{equation}
where $L^\lambda{}_{\mu\nu}$ is the disformation tensor.
It follows that,
\begin{equation} \label{L}
L^\lambda{}_{\mu\nu} = \frac{1}{2} (Q^\lambda{}_{\mu\nu} - Q_\mu{}^\lambda{}_\nu - Q_\nu{}^\lambda{}_\mu) \,.
\end{equation}
We can construct two different types of non-metricity vectors,
\begin{equation*}
 Q_\mu := g^{\nu\lambda}Q_{\mu\nu\lambda} = Q_\mu{}^\nu{}_\nu \,, \qquad \tilde{Q}_\mu := g^{\nu\lambda}Q_{\nu\mu\lambda} = Q_{\nu\mu}{}^\nu \,.
\end{equation*}
The superpotential (or the non-metricity conjugate) tensor $P^\lambda{}_{\mu\nu}$ is given by
\begin{equation} \label{P}
P^\lambda{}_{\mu\nu} = \frac{1}{4} \left( -2 L^\lambda{}_{\mu\nu} + Q^\lambda g_{\mu\nu} - \tilde{Q}^\lambda g_{\mu\nu} -\frac{1}{2} \delta^\lambda_\mu Q_{\nu} - \frac{1}{2} \delta^\lambda_\nu Q_{\mu} \right) \,.
\end{equation}
Finally, the non-metricity scalar $Q$ is defined as
\begin{equation} \label{Q}
Q=Q_{\alpha\beta\gamma}P^{\alpha\beta\gamma}.
\end{equation}
It should be mentioned that $R^\rho_{\sigma \mu \nu} = 0$ is one of the restrictions that was utilised when developing the $f(Q)$-theory. This indicates that there is a unique coordinate system that can be chosen to make the affine connection disappears, denoted by the expression $\Gamma^\lambda_{\mu\nu} = 0$. The term ``coincident gauge" refers to this kind of circumstance. In this particular setting, the field equation of the affine connection is trivially satisfied by all $f(Q)$ models. So we only concentrate on the field equation obtained by varying the action phrase
\begin{equation} \label{action}
S = \int \left[\frac{1}{2}f(Q) + \mathcal{L}_M \right] \sqrt{-g}\,d^4 x
\end{equation}
with respect to the metric \cite{coincident}
\begin{equation} \label{FE1}
\frac{2}{\sqrt{-g}} \nabla_\lambda (\sqrt{-g}f_QP^\lambda{}_{\mu\nu}) -\frac{1}{2}f g_{\mu\nu} + f_Q(P_{\nu\rho\sigma} Q_\mu{}^{\rho\sigma} -2P_{\rho\sigma\mu}Q^{\rho\sigma}{}_\nu) = T^m_{\mu\nu}.
\end{equation}
$T^m_{\mu\nu}$ is the energy-momentum tensor generated from the matter Lagrangian $\mathcal{L}_M$, which additionally is assumed to be independent of the affine connection. We assume a barotropic perfect fluid including dark matter $T^m_{\mu\nu}=(p+\rho)u_\mu u_\nu+pg_{\mu\nu}$ with isotropic pressure $p$, energy-density $\rho$ and the four-velocity vector $u^\mu$. Thereafter, using the relation (\ref{connc}), after some calculations (see \cite{zhao} and the Appendix of \cite{gde}) we can rewrite the field equations (\ref{FE1}) in its covariant formulation as
\begin{equation} \label{FE2}
f_Q \mathring{G}_{\mu\nu} + \frac{1}{2}g_{\mu\nu}(Qf_Q - f) + 2f_{QQ} \mathring{\nabla}_\lambda Q P^\lambda{}_{\mu\nu} = T^m_{\mu\nu}
\end{equation}
where $$\mathring{G}_{\mu\nu} = \mathring{R}_{\mu\nu} - \frac{1}{2} g_{\mu\nu} \mathring{R}.$$ All the expressions with a $\mathring{()}$ is calculated with respect to the Levi-Civita Connection $\mathring{\Gamma}$. Hence in its GTR equivalent effective form we can express it as
\begin{align}\label{eff}
    \mathring{G}_{\mu\nu}=\frac{1}{f_Q}T^{eff}_{\mu\nu}=\frac{1}{f_Q}T^m_{\mu\nu}+T^{DE}_{\mu\nu},
\end{align}
where
\begin{align*}
T^{DE}_{\mu\nu}=\frac{1}{f_Q}\left[\frac{1}{2}g_{\mu\nu}(f-Qf_Q)-2f_{QQ}\mathring{\nabla}_\lambda QP^\lambda_{\mu\nu}\right].
\end{align*}

\section{\textbf{Field equations and some consequences:}}\label{sec2}
In the current discussion, we consider the axially symmetric Bianchi universe with line element given in terms of Cartesian coordinates
\begin{equation}\label{metric}
ds^2 = -\mathrm{d} t\otimes \mathrm{d} t + e^{\left(-4 \, \chi\left(t\right) + 2 \, \xi\left(t\right)\right)} \mathrm{d} x\otimes \mathrm{d} x + e^{\left(2 \, \chi\left(t\right) + 2 \, \xi\left(t\right)\right)} \mathrm{d} y\otimes \mathrm{d} y + e^{\left(2 \, \chi\left(t\right) + 2 \, \xi\left(t\right)\right)} \mathrm{d} z\otimes \mathrm{d} z.
\end{equation}
Let us mention that the above locally rotationally Bianchi-1 metric has a unique feature. For example, starting from the most general linearly perturbed spatially flat FLRW metric
\begin{equation}   g = -(1+2\Phi(\bar{x},t))dt^2-2B_i(\bar{x},t)dtdx^i+a^2(t)[(1-2\Psi(\bar{x},t))\delta_{ij}+2h_{ij}(\bar{x},t)]dx^idx^j,
\end{equation}
under synchronous gauge ($\Phi = 0, \,B_i = 0$) and the homogeneity bound ($\Psi,\,h_{ij}$ function of time alone), it is possible to obtain such an anisotropic metric \eqref{metric} \cite{saikat}. This bridges the gap between the isotropic and homogeneous FLRW metric and the homogeneous Bianchi type metric with very small anisotropy as considered here. The Ricci scalar $\mathring{R}$ and the non-metricity scalar $Q$ are calculated to be

\begin{equation} \label{R}
\mathring{R}= 6 \dot{\chi}^2 + 12 \dot{\xi}^{2} + 6 \ddot{\xi};~~~~~Q(t)=6\dot{\chi}^2-6\dot{\xi}^2.
\end{equation}
The equations of motion are:

\be\label{1}\begin{split}  &{f\over 2} - Qf_Q = \rho,\\&
{f\over 2} + 2f_Q(\ddot \xi + \ddot\chi + 3\dot \xi^2 + 3\dot\xi\dot\chi) + 2\dot Q f_{QQ}(\dot\xi+\dot\chi) = - p,\\&
{f\over 2} + 2f_Q(\ddot \xi - {1\over 2} \ddot\chi + 3\dot \xi^2 - {3\over 2}\dot\xi\dot\chi) + 2\dot Q f_{QQ}(\dot\xi-{1\over 2}\dot\chi) = -p.\end{split}\ee
Here $\rho$ and $p$ stand for the energy density and the thermodynamic pressure of a perfect barotropic fluid. The non-metricity scalar, the effective average Hubble parameter and the effective average scale factors may be expressed as,

\be\begin{split} \label{3} & Q = -6(\dot\xi^2 - \dot\chi^2) = {2\over 3}(3\sigma^2-\theta^2) = 2(\sigma^2 - 3H^2),\\&
\theta = 3H = 3\dot \xi,~~ a = e^\xi,~~\sigma^2 = 3\dot \chi^2.\end{split}\ee
where $\theta$ is the expansion scalar, $\sigma^2$ the shear scalar, while the proper volume is $\mathcal{V}=e^{3\xi}$.\\

\noindent
From the pair of pressure equations one immediately arrives at,
\be \label{4} f_Q\dot \chi = c_1 e^{-3\xi}\ee
and it produces
\be\label{fQ} \dot Q f_{QQ} = - f_Q\left[3 \dot\xi + {\ddot \chi\over \dot \chi}\right].\ee
In view of equations \eqref{1} and \eqref{fQ}, one can also construct the following equation,

\be\label{rhop} \rho - 3p = 2 f + 6f_Q\left(\ddot\xi + \dot\xi^2 - \dot\chi^2 - \dot \xi{\ddot\chi\over \dot \chi}\right),\ee
which we shall use shortly. The covariant conservation of the energy momentum in the present scenario follows from \cite{de/epjc}. This checkpoint is not trivial as in the case of GTR or $f(R)$ theories. Since in certain scenarios, for example, in a static spherically symmetric spacetime, this imposes one additional constraint over the model parameters. One can show that the covariant conservation of energy momentum is equivalent to the field equation obtained by varying the same $f(Q)$ action term with respect to the affine connection (recall that the symmetric teleparallel theory is necessarily a metric-affine theory), but the situation does not improve any further, the model parameters has to satisfy this second field equation now \cite{ad/bianchi}. In the present scenario, however, we do not have any additional burden arising from this. The result can also be shown directly from the equations of motion (\ref{1}). Differentiating the energy equation of (\ref{1}) with respect to time variable, we obtain

\begin{eqnarray} \dot{\rho}&=&-\frac{\dot{Q}}{2}f_Q-Q\dot{Q}f_{QQ}\notag\\
&&=f_Q\left[3\dot{\xi}Q+\frac{\ddot{\chi}}{\dot{\chi}}Q-\frac{\dot{Q}}{2}\right], \text{ due to  } (\ref{fQ})\notag\\
&&=6f_Q\left[3(\dot{\chi}^2-\dot{\xi}^2)+\ddot{\xi}-\dot\xi\frac{\ddot{\chi}}{\dot{\chi}}  \right]\dot{\xi}.
\end{eqnarray}
On the other hand, using (\ref{fQ}) in (\ref{1}) we also have
\begin{eqnarray}
    p+\rho=-2f_Q\left[3(\dot{\chi}^2-\dot{\xi}^2)+\ddot{\xi}-\dot\xi\frac{\ddot{\chi}}{\dot{\chi}}  \right]
\end{eqnarray}
which combined to yield the usual continuity relation $\dot{\rho}=-3\dot{\xi}(p+\rho)$.
Hence, one can instead use the set of three independent equations (the density equation, one of the pressure equations and the relation \eqref{4}) with five ($f(Q),\rho,p,\xi,\chi$) unknowns. Clearly we need two physically legitimate assumptions to obtain exact solutions. Other than the barotropic equation of state $p = \omega \rho$, the most legitimate assumption is $\sigma^2 \propto \theta^2$, which has been widely used earlier in different Bianchi models \cite{B1,B2,B3,B4,B5,B6,B7}.
In Appendix \ref{appI}, we show that in radiation era, a quadratic model $f(Q)=f_0Q^2$ reproduces the condition $\sigma^2 \propto \theta^2$ in the current scenario.
Additionally, it has been found that if a relation between the metric coefficients are assumed in orthogonal, locally rotationally-symmetric and spatially-homogeneous Bianchi II, VIII and IX metrics, the relation $\sigma^2 \propto \theta^2$ automatically follows \cite{B8}. Finally, de-Sitter solution in modified theory of gravity associated with higher order curvature invariant terms, also leads to the same relation $\sigma^2 \propto \theta^2$ \cite{B9, B10}. However under such assumption, here one obtains,

\be\label{5}\theta^2 \propto \sigma^2,~~ \mathrm{i.e.,}~~~\dot\xi = k\dot\chi.\ee
As a result,
\be\label{5a} \xi = k\chi + k_1, ~~\mathrm{and}~~ Q = -6(k^2-1)\dot\chi^2 = -6\dot\xi^2\left({{k^2-1}\over k^2}\right).\ee
Now, in view of equation \eqref{rhop}, one finds,

\be\label{rhop1} \rho -3p = 2f + 6\dot\xi^2\left({{k^2-1}\over k^2}\right) f_Q = 2f - Q f_Q.\ee
\subsection{Limitation and remedy}
\begin{itemize}
    \item From (\ref{rhop1}), it is clear that in the radiation era $p = {1\over 3}\rho$, $f(Q) = f_0 Q^2$ is the only allowed form for $f(Q)$, and `General theory of Relativity' can not be retrieved in any way. 
    \item Substituting \eqref{5} and \eqref{5a} in the pressure equation, one obtains $f(Q) = 0$, both in the pressure-less dust era ($p=0$) as well as in the early vacuum dominated era ($p = \rho = 0$). 
    \item In the case of pure vacuum ($\rho = p = 0$), the first equation of the set \eqref{1} yields $f(Q) \propto \sqrt Q$.
\end{itemize} 

 Note that, a non-negative $Q$ warrants the expansion rate is smaller than the shear scalar ($\theta^2 < 3\sigma^2$) throughout the cosmic evolution, and the universe never isotropizes. In this context, we recall that such uncanny behaviour is absent from $f(R)$ theories of gravity, which admits de-Sitter solution in pure vacuum, and therefore has much rich structure than $f(Q)$ gravity. However, in order to avert such pathologies, we need either to withdraw the assumption \eqref{5} or to modify the right hand side of the field equations. As already mentioned, the assumption \eqref{5} has largely been used in the literature earlier, and also appears automatically in some cases, so the simplest modifications are either to associate viscosity and/or a scalar field. In the present work, we consider a scalar field $\phi$, which is required to drive inflation in the very early universe. Note that this is not a Quintessence field which is required to explain late-time cosmic evolution with accelerated expansion followed by deceleration. This scalar field happens to be present in the very early universe, which may be a Higgs field or a dilatonic one. Such a scalar field not only drives inflation, but also puts an end to it, and happens to be responsible for particle creation, as it oscillates after the end of inflation. This field must have been redshifted largely and should not exist today, or if exists, is almost undetectable. In view of such a scalar field, the independent set of field equations are therefore,
\be\label{6}\begin{split}  &{f\over 2} - Qf_Q = \rho +{1\over 2}\dot\phi^2 + V(\phi),\\&
{f\over 2} + 2f_Q(\ddot \xi + \ddot\chi + 3\dot \xi^2 + 3\dot\xi\dot\chi) + 2\dot Q f_{QQ}(\dot\xi+\dot\chi) = - p - {1\over 2}\dot\phi^2 + V,\\&
f_Q\dot \chi=c_1e^{-3\xi}.\end{split}\ee
Additionally, we have the Bianchi identity and the Klein-Gordon equation, which are,
\be\label{7}\begin{split}  &\dot \rho + 3\dot\xi(\rho+p) = 0,\\&
\ddot\phi +3\dot\xi\dot\phi + V'(\phi) = 0.\end{split}\ee
Combining, we also find

\be\label{rhop2} \rho - 3p - \dot\phi^2 + 4V = 2f + 6f_Q\left[\ddot\xi + \dot\xi^2 - \dot\xi{\ddot\chi\over\dot\chi} - \dot\chi^2\right].\ee

\subsection{Energy condition:}

Before we proceed further, it is necessary to fix the signature of the coefficients $(\alpha~\mathrm{and}~ \beta)$ associated with the model $f(Q) = \alpha Q + \beta Q^2$, and its subcase $f(Q) = \beta Q^2$ that we consider to study inflation and else. This is possible in view of energy conditions. We remind that the essence of all the modified theories of gravity including the symmetric telleparallel geometry under present consideration, developed so far, is to resolve the cosmic puzzle without the need of dark energy. Thus, although scalar field has been considered to study inflation, it must not remain present at the end. Hence much of the scalar field should be redshifted during radiation era, and all we shall left with is the barotropic fluid in the pressureless dust era. In addition, the universe should turn out to be isotropic, as par current observation, hence contribution from shear should be considerably low. For perfect fluid $T_{\mu\nu} = (\rho + p)u_\mu u_\nu + p g_{\mu\nu}$, the energy conditions are:\\
1. Null energy condition: $\rho + p \ge 0$.\\
2. Weak energy condition: $\rho + p \ge 0$ and $\rho \ge 0$.\\
3. Dominant energy condition:$\rho \ge |p|$.\\
4. Strong energy condition:$\rho + p \ge 0$ and $\rho + 3p \ge 0$.\\
It is important to mention that $p < 0$ is allowed, and strong energy condition does not automatically satisfy the weak energy condition. However, if both $\rho > 0$, and $p > 0$, all energy conditions are satisfied simultaneously.\\

For the present purpose, equations \eqref{6} may be cast in the following manner:
\be\begin{split}\label{EC}& \rho = {f\over 2} - Qf_Q -\rho_\phi,\\&
\rho+p = -Qf_Q - 2c_1 e^{-3\xi}{d\over dt}\left(H\over\sigma\right) - (\rho_\phi + p_\phi),\\&
\rho+3p = -f(Q) - Qf_Q - 6c_1 e^{-3\xi}{d\over dt}\left(H\over\sigma\right) - (\rho_\phi + 3p_\phi).\end{split}\ee
In the above, $\rho_\phi = {1\over 2}\dot \phi^2 + V(\phi)$, and $p_\phi = {1\over 2}\dot \phi^2 - V(\phi)$. Since in the matter dominated era, $\dot\xi = H > 0$ while, $\dot\chi = \sigma$ may be positive or negative, if not zero, the sign of the second term appearing on the right hand side of  both the second and the third equations of \eqref{EC} remains arbitrary. However, as the universe is large enough ($e^{3\xi} = a^{3}$), this term contributes little, and as long as the assumption \eqref{5} holds, the term vanishes.\\

1. Now for `General theory of Relativity', $f(Q) = \alpha Q = \alpha(2\sigma^2-6 H^2)$. Hence, $\rho = -{\alpha Q \over 2} -\rho_\phi$, $\rho+p = -\alpha Q - 2c_1 e^{-3\xi}{d\over dt}\left(H\over\sigma\right) - (\rho_\phi + p_\phi)$ and $\rho+3p = -2\alpha Q - 6c_1 e^{-3\xi}{d\over dt}\left(H\over\sigma\right) - (\rho_\phi + 3p_\phi)$. Hence, only if anisotropy is small, $\sigma^2 < 3H^2$, then $Q < 0$, and all the energy conditions may be satisfied for $\alpha > 0$ ($\alpha = {1\over 16\pi G}$, say in the unit $\hbar = c =1$), and contribution from the scalar field, in particular, is small.\\

2. Next let us consider, $f(Q) = \beta Q^2$. Clearly, $\rho = -{3\over 2}\beta Q^2 -\rho_\phi$ indicates $\beta < 0$. Further, $\rho + p = -2\beta Q^2 - 2c_1 e^{-3\xi}{d\over dt}\left(H\over\sigma\right) -(\rho_\phi + p_\phi) > 0$, and $\rho + 3p = -3\beta Q^2 - 6c_1 e^{-3\xi}{d\over dt}\left(H\over\sigma\right) -(\rho_\phi + 3p_\phi) > 0$ also require $\beta < 0$, and little contribution from the scalar field. \\

3. Finally, for $f(Q) =  \alpha Q + \beta Q^2$, $\rho = -{\alpha\over 2} Q - {3\over 2} \beta Q^2 -\rho_\phi$, $\rho + p = -\alpha Q -2\beta Q^2 - 2c_1 e^{-3\xi}{d\over dt}\left(H\over\sigma\right) -(\rho_\phi + p_\phi)$ and $\rho + 3p = -2\alpha Q - 3\beta Q^2 - 6c_1 e^{-3\xi}{d\over dt}\left(H\over\sigma\right) -(\rho_\phi + 3p_\phi)$. Clearly again, our earlier choice, $\alpha > 0$ and $\beta < 0$ ensure $\rho > 0$, throughout cosmic evolution, provided contribution from the scalar field remains subdominant.\\

In a nutshell, energy conditions in all the three cases are satisfied provided, $\alpha > 0$, $\beta < 0$ and $|\alpha Q| > 2\rho_\phi$, $|\alpha Q| > \rho_\phi + p_\phi$ hold. It is worth mentioning that the sign of $c_1$ may be fixed from the solution of the field equations. \\

Let us emphasize the fact that, if most of the scalar is used up in creating particles, at the end of inflation and during re-heating, while the rest is redshifted in the following radiation dominated era, by some means, then matter-dominated era, would be left without the scalar field. Field equations for $f(Q) = \alpha Q - \beta Q^2$, then can be formulated in terms of the effective energy density and the effective pressure in the following forms:

\be \label{preff}\begin{split}& -{\alpha\over 2}Q = \rho - {3\over 2}\beta Q^2 = \rho_e,\\&
{\alpha\over 2}Q  + 2\alpha \dot\chi {d\over dt}\left({\dot\xi\over \dot\chi}\right) = - {1\over 2}\beta Q^2 - 2\beta Q\dot\chi{d\over dt}\left({\dot\xi\over \dot\chi}\right) = -p_e,\end{split}\ee
so that, the effective equation of state $\omega_e$ reads as:

\be\label{eos} \omega_e = {p_e\over \rho_e} = {{1\over 2}\beta Q^2 + 2\beta Q\dot\chi{d\over dt}\left({\dot\xi\over \dot\chi}\right)\over \rho - {3\over 2}\beta Q^2}.\ee
Since, the second term in the numerator may possibly be negative, so early deceleration followed by acceleration in the matter dominated era may be envisaged.

\subsection{In search of vacuum de-Sitter solution:}

Let us now seek de-Sitter solution in the form $a = e^\xi = a_0 e^{\Lambda t},~ \mathrm{i.e.},~  \dot \xi = \Lambda$, under the assumption,
\be\label{st}\sigma^2 \propto \theta^2, ~\mathrm{for~ which},~ \dot \xi = k\dot \chi.\ee
Note that in the case of $f(R)$ gravity, de-Sitter solution enforces a constant Ricci scalar. Likewise, under the above choice \eqref{st}, $Q = - {6(k^2 - 1)\over k^2}\Lambda^2$ also becomes a constant. In the vacuum dominated era ($p = 0 = \rho$), the set of field equations \eqref{6} and \eqref{7} simplify to:

\be\label{8} \begin{split}
&{f\over 2} - Q f_Q = {1\over 2} \dot\phi^2 + V(\phi),\\&
{f\over 2} = - {1\over 2} \dot\phi^2 + V(\phi),\\&
f_Q\dot\chi = c_1 e^{-3\xi},\\&
\ddot \phi + 3\dot\xi\dot\phi + V'(\phi) = 0.\end{split}\ee
Note that the Klein-Gordon equation is not an independent one, and so we now have three independent equations with ($f(Q), \chi, \phi, V(\phi)$) four unknowns. From the first pair we find,

\be \label{9}(Qf_Q) \dot\chi = -\dot\phi^2\dot\chi, ~~\mathrm{or,}~~\dot\phi^2 = 6c_1(k^2 - 1)\dot\chi e^{-3\xi}, ~~\mathrm{or,}~~ \dot\phi^2 = 6c_1\left({k^2 - 1\over k}\right)\dot\xi e^{-3\xi},\ee
in view of the assumption \eqref{st}. Now as de-Sitter solution is imposed, we find,

\be\label{11}\dot\phi^2= 6c_1\Lambda\left({ {k^2-1}\over k a_0^3}\right)e^{-3\Lambda t}.\ee
At this point, we recall that the convention adopted in the present study leads to $Q = -6H^2 < 0$ in homogeneous and isotropic Robertson-Walker space-time. Hence, if shear is not too large from the beginning, $3\sigma^2 < \theta^2$, implying $k > 1$, and so $c_1 > 0$. Now, \eqref{11} may be integrated to yield

\be\label{12} \phi = \left[\phi_0 - {2\over 3\Lambda}\sqrt{{6\Lambda c_1(k^2-1)\over k a_0^3}}e^{-{3\over 2}\Lambda t}\right].\ee
Now under repeated differentiation of \eqref{12} we find,

\be \label{13} \dot\phi = {3\Lambda\over 2}(\phi_0-\phi),~~~\mathrm{and}~~~\ddot \phi 
= -{9\over 4}\Lambda^2(\phi_0-\phi). \ee
Therefore we can find the form of the potential from the Klein-Gordon equation as:

\be \label{14} V' = -{9\over 4}\Lambda^2 (\phi - \phi_0),~\mathrm{and~so}~~ V(\phi) = -{9\over 8}\Lambda^2\phi^2 + {9\over 4}\Lambda^2\phi_0\phi + V_0.\ee
Unfortunately,
\be \label{15} f(Q) = -{9\over 4}\Lambda^2\phi_0^2-{9\over 2}\Lambda^2\phi^2 + 9\Lambda^2\phi_0\phi+ 2V_0,\ee
cannot be expressed in terms of $Q$ alone. In any case there appears a contradiction, since for $Q = $constant, $f(Q)$ must also be a constant, while here $f(Q)$ evolves with time along with the scalar field $\phi$. In a host of modified gravity models, de-Sitter solution is admissible under the choice $\phi = \phi_0 e^{-\Lambda t}$ in isotropic metric \cite{HO1,HO2,HO3,HO4,HO5,HO6,HO7,HO8}, as well as in anisotropic Bianchi metrics \cite{B9,B10}. However, such assumption here leads to imaginary $\chi(t)$. In a nutshell, a legitimate de-Sitter solution is not admissible in $f(Q)$ theory of gravity. This is definitely yet another serious pathology of the theory.\\

Nonetheless, we can still study inflation with exponential expansion, since while studying inflation, our assumption of de-Sitter solution is relaxed, and $\dot \xi$ is assumed to vary slowly, instead of being a constant ($\Lambda$). Thus we can choose arbitrary form of the potential and $f(Q)$. Further slow roll inflation requires a flat potential, which we shall choose for the purpose.

\section{Scalar field inflation}\label{sec3}

Despite some alternative suggestions \cite{alt1, alt2, alt3}, Inflation has been contemplated as a scenario, since it is the most viable model that can resolve the issues like flatness, horizon, monopole and structure formation problems singlehandedly, which were left unanswered by FLRW model. Recent analysis of the combined data (TT, TE, EE + lowE +lensing + BK15 + BAO+ Bicep2) puts up tight constrains on inflationary parameters, viz., the scalar to tensor ratio $(r < 0.055)$ and (scalar) spectral index $(0.9631 < n_s < 0.9705)$, along with a very small tensor spectral index $n_t$ {\cite{Planck1},\cite{Planck2}}. It is already known that to solve the horizon and the flatness problems, the number of e-folds $N$ should preferably lie within the range $45 < N < 65$. In fact, a more recent analysis \cite{Planck3} suggested even tighter constraint on scalar to tensor ratio $(r < 0.032)$. It's true that the constraints on the inflationary parameters are model dependent. Nonetheless, as mentioned, constraints are put forward in view of a combined data set along with a totally different analysis. It is therefore suggestive that theoretical models  should admit these observational constraints at least to some extent. In this section we proceed with the said motivation. For this purpose, we consider a vacuum-dominated $(p_m =0 ;~~ \rho_m=0)$ scalar field driven inflation in the very early universe. To check if our model can overcome such observational constraints, we adopt the following methodology. First, we simplify the field equations considerably, applying slow-roll assumption and compute the inflationary parameters $\epsilon,~\eta, N(\phi)$. Next we choose an initial value of the scalar field, and try to find a good fit following trial and error method. Particularly, since initially $\epsilon_i = \frac{r}{16}$, so we choose $\epsilon_i$ according to the experimental constraint, taking $k^2 > 1$, required to keep $Q < 0$. Then, since inflation halts at $\epsilon_f = 1$, so it allows us to choose a ratio of $\phi_i$ and $\phi_f$, and hence $\phi_f$ is fixed. Next, we choose $\beta V_1 = 1$, and select $\frac{V_0}{V_1}$  by trial and error method, so that $n_s$ fits Planck's constraint. Since it yields $N$ within the limit, so we conclude that the fit is perfect with recently released Planck's data. Finally, we find that inflation is sub-Planckian, and the universe comes out of this inflationary era, being determined by the oscillatory behaviour of the scalar field. Here we present three sets of data. There may be other sets of data too, choosing different ratio of $\phi_i$  and $\phi_f$, $\beta V_1$, and $\frac{V_0}{V_1}$. We start from the effective Friedmann equation, that may be written as,

\begin{align}
\left[{\dot\phi^2\over 2}+V(\phi)-\frac{f}{2}+Qf_Q\right]=0,\label{eq33}\end{align}
and the Klein Gordon equation can be written in its standard form as,
\begin {align} \ddot\phi +3H{\dot\phi}+ V'(\phi)=0 .\label{KG}\end{align}

\subsection{$f(Q) = -\beta Q^2,~\beta > 0$:}

Since we have found $f(Q)\propto Q^2$ in radiation era, so we first check its outcome in inflationary regime. Under the above quadratic form of $f(Q)$, the effective Friedmann equation \eqref{eq33} reads as,
\begin{align} -54\beta\frac{(k^2-1)^2}{k^4}H^4+\left[{\dot\phi^2\over 2}+V(\phi)\right]=0.\label{Fried}\end{align}
Now, enforcing the standard slow-roll conditions $\dot\phi^2\ll V(\phi)$ and $|\ddot\phi|\ll 3{H}|\dot\phi|$, equations (\ref{Fried}) and (\ref{KG}), reduce to,
\begin{align} -54\beta \frac{(k^2-1)^2}{k^4}H^4+V(\phi)=0,~~~~~~3{H}\dot\phi \simeq - V'.\label{com}\end{align}
Solving $H^2$ from \eqref{com} one obtains,
\begin{align}H^2={\frac{k^2}{\sqrt{54}(k^2-1)}\sqrt {V(\phi)\over \beta}}.\label{soln}\end{align}
Now, combining the pair of equations (\ref{com}), one can show that the potential slow roll parameter $\epsilon$ equals the Hubble slow roll parameter ($\epsilon_1$) under the condition,
\begin {align}\label{SR}\begin{split} \epsilon \equiv - {\dot {H}\over {H}^2}=\frac{\sqrt{3\beta}(k^2-1)V'^2}{2{\sqrt 2}k^2V^{3\over 2}};\hspace{0.3 in}\eta = 2 \alpha \left({V''\over V}\right);\end{split}\end {align}
where $\epsilon \ll 1$, during inflation. The primordial spectral index of scalar perturbation $(n_s)$, the tensor-to-scalar ratio $(r)$, and the tensor spectral index $(n_t)$ may now be expressed as,
\begin{align} n_s = 1 - 6\epsilon + 2\eta,~~~~ r = 16\epsilon, ~~~ n_t = -{r\over 8}=-2\epsilon,\end{align}
where, $|n_s| < 1$ ensures deviation from scale invariance. Further, since $\frac{H}{\dot\phi}=-\frac{k^2\sqrt{ V(\phi)}}{\sqrt{6\beta}(k^2-1)V'}$, therefore, the number of e-folds  may be computed as follows:

\begin{align}\label{Nphi} {N}(\phi)\simeq \int_{t_i}^{t_f}{H}dt=\int_{\phi_i}^{\phi_f}\frac{{H}}{\dot\phi}d\phi\simeq \int_{\phi_f}^{\phi_i}\Big{(}\frac{k^2\sqrt{ V(\phi)}}{\sqrt{6\beta}(k^2-1)V'}\Big{)}d\phi,\end{align}
where, $\phi_i$ and $\phi_f$ denote the scalar fields at the beginning $(t_i)$ and at the end $(t_f)$ of inflation. Let us mention that, Albrecht and Steinhardt \cite{AS} proposed an important gravitational effect called the Hubble friction, which sustains inflation for sufficiently long period. This is referred to as `slow roll inflation'. It is noteworthy that Hubble friction is required for generating a spectrum of density fluctuations at the end of inflation, that results in structure formation. We therefore choose a potential that has a plateau, so that it slowly rolls and inflation persists for sufficiently long period, and then stops as the plateau is traversed. Let us therefore choose the potential in the following form,

\begin{align}
V(\phi)= V_0-{V_1\over \phi},
\end{align}
so that the potential remains almost flat as $\phi$ is large, and slow roll is admissible. There may exist other variety of potential, however, it is beyond the scope of present work to include all. Now, as $V(\phi)= V_0-{V_1\over \phi}$,~~~ $V'(\phi)= {V_1\over \phi^2}$ and $V''(\phi)=-{2V_1\over \phi^3}$, then the equations
\eqref{SR} and \eqref{Nphi} reduces to,
\begin {align}\label{para}\begin{split} \epsilon =\frac{\sqrt{3\beta V_1}(k^2-1)}{2{\sqrt 2}k^2\phi^{5\over 2}(\frac{V_0}{V_1}\phi-1)^{3\over 2}},\hspace{0.5in}\eta=-\frac{2M_P^2}{\phi^2(\frac{V_0}{V_1}\phi-1)},\hspace{0.5in} N=\int_{\phi_f}^{\phi_i}\Big{[}\frac{k^2\phi^{3\over 2} (\frac{V_0}{V_1}\phi-1)^{1\over 2}}{\sqrt{6\beta V_1}(k^2-1)}\Big{]}d\phi.\end{split}\end {align}

\begin{figure}
\begin{minipage}[h]{0.47\textwidth}
      \centering
      \begin{tabular}{|c|c|c|c|}
     \hline\hline
      $\phi_i$ in $M_P$  & $n_s$ & $r$ & $ {N}$\\
      \hline
       4.00 &  0.96277 & 0.00403 & 45\\
       4.05 &  0.96422 & 0.00383 & 47\\
       4.10 &  0.96559 & 0.00363 &49\\
       4.15 &  0.96689 & 0.00345 & 51\\
       4.20 &  0.96813 & 0.00328 & 53\\
       4.25 &  0.96930 & 0.00312 & 55\\
       4.30 &  0.97042 & 0.00297 & 57\\
       4.35 &  0.97148 & 0.00283 & 59\\
    \hline\hline
    \end{tabular}
      \captionof{table}{Data set for the inflationary parameters with $\phi_f=0.702211~M_P$, $\beta V_1=1~M_P^5$,~${V_0\over V_1}=2~M_P^{-1}$ and $k=1.15$, varying $\phi_i$.}
      \label{tab:Table1}
   \end{minipage}
   \hfill%
  \begin{minipage}[h]{0.47\textwidth}
\includegraphics[ width=1.10\textwidth] {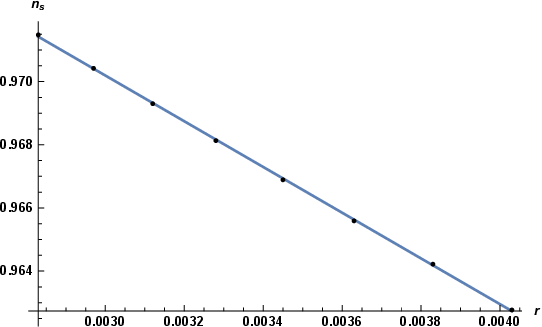}
 \caption{Variation of $n_s$ with $r$.}
      \label{fig:1}
   \end{minipage}
\end{figure}

\noindent
In Table 1, we present a data set for the inflationary parameters, varying $\phi_i$ between $4.00M_P \leq\phi_i\leq 4.35M_P$, so that $r$ and $n_s$ lie within the experimental limit. Note that slow rollover terminates $(\epsilon = 1)$ at $\phi_ f = 0.702211M_P$ in every case. Further, the number of e-folds $45 \leq N \leq 59$ found is sufficient to solve the horizon and flatness problems. The data corresponding to inflationary parameters ($r <0.0041$, and $0.962 < n_s < 0.971$) are perfectly at par with current observation\cite {Planck1,Planck2}. The variation of spectral index $n_s$ with the scalar to tensor ratio $r$ is shown in figure 1. Let us now proceed to find the energy scale of inflation in view of the relation \eqref{soln}. Taking into consideration the data set of Table 1, associated with $N = 57$, for which $\phi_i = 4.30M_P$, we find
\begin{align}{H_*}^2=\frac{1.3225{\sqrt{\beta V_1\left(\frac{V_0}{V_1}-\frac{1}{\phi_i}\right)}}}{{\sqrt54}(1.3225-1)\beta}=\frac{1.75826}{2.3699\beta}=\frac{0.74191}{\beta}, ~~~~\text{or}~~~~~~~~~~~~~~~~ H_*=\sqrt{\frac{0.77131}{\beta}},\label{42} \end {align}
where we have considered $k =1.15 $, ${V_0\over V_1}= 2 M_P^{-1}$ and $\beta V_1 = 1M_P^5$. Note that $\beta$ still remains arbitrary. Now, the energy scale of inflation in a single scalar field model corresponding to GTR is given by the expression \cite{Wands} as,

\begin{align}{H_*} = 8\times 10^{13}\sqrt{\frac{r}{0.2}}GeV = 9.76 \times 10^{12} GeV \approx 3.93\times 10^{-6}M_P.\label{43}\end{align}
In the above computation, we have used the data $r = 0.00297$ of Table 1. Thus, in order to match the scale of inflation \eqref{42}  with the single field scale of inflation \eqref{43}, $\beta$ has to be of the order, $\beta\approx 4.80 \times 10^{10} $. Note that, we have fixed only two parameters $\beta V_1$ and $V_0\over V_1$, for the present inflationary model. However, with the above consideration (sub-Planckian scale of inflation), the parameter $\beta$ is fixed. Further, since individually each term has to be of the same order of magnitude, one can compare the two terms of \eqref{soln} to find that:
\begin{align} V(\phi) = V_1\left({V_0\over V_1}-{1\over \phi}\right)=1.76V_1= 3.678\times 10^{-11}M_P^4 ,\end{align}
implying \begin{align}\label{V1} V_1\approx 2.09\times10^{-11}M_P^5, ~~~~~~V_0 \approx 4.18\times 10^{-11}M_P^4.\end{align}
Thus, all the parameters are fixed once and forever.\\

Finally, we need to test whether the model gracefully exits from inflation. Thus, with the form of the potential, $V(\phi)=(V_0-{V_1\over \phi})$, with $\beta > 0$, equation \eqref{Fried} may be expressed as,

 \begin{align}\label{osc}{54 (k^2-1)^2H^4\over k^4V_1}= \frac{1}{\beta}\left[{\dot\phi^2\over 2V_1}+{V_0\over V_1}-{1\over\phi}\right].\end{align}
During inflation, $H^2$ and $V_1$ are of the same order of magnitude, while Hubble parameter varies slowly. However, at the end, Hubble rate usually decreases sharply, and  $H^4$ falls much below $V_1$. Hence, $ H^4\over V_1$ may be neglected without loss of generality. Thus, one can write
\begin{align}{\dot{\phi}}^2=-2\left[V_0-{V_1\over \phi}\right].\label{osc1}\end{align}
Let us choose the following oscillatory form of $\phi$ to check if the consequence is physically admissible.

\begin{align}\phi = \exp({i\omega t}).\end{align}
Therefore, equation \eqref{osc1} reduces to,

\begin{align} \omega^2\phi^2=2\left(V_0-{V_1\over \phi}\right).\end{align}
Hence, with $V_0=4.18\times 10^{-11}M_P^4,~~V_1=2.09\times 10^{-11}M_P^5,~\phi_f=0.702211M_P$, one finds $\omega= 6.98\times 10^{-6}M_P$.
Clearly, $\phi$ exhibits oscillatory behaviour, provided $\omega= 6.98\times 10^{-6}M_P$ .

\subsection{$f=\alpha Q-\beta Q^2,\alpha>0,~\beta>0:$}
 In this case, for vacuum era $ ( p_m =0 ;~~ \rho_m=0)$ the Friedmann equation may be expressed as,

\begin{align}  {\dot\phi^2\over 2}+V(\phi)-\left(\frac{\alpha Q-\beta Q^2}{2}\right)+(\alpha-2\beta Q)Q=0.\end{align}
Again using the equation \eqref{5} the above equation reduces to,

\be\label{FM2} -54\beta H^4\left({{k^2-1}\over k^2}\right)^2-3\alpha H^2\left({{k^2-1}\over k^2}\right)+\left[{\dot\phi^2\over 2}+V(\phi)\right]=0,\ee
and the Klein Gordon equation can be written in its regular form as,

\begin {align} \ddot\phi +3H{\dot\phi}+ V'(\phi)=0.\label{KG2}\end{align}
We now enforce the standard slow-roll conditions $\dot\phi^2\ll V(\phi)$ and $|\ddot\phi|\ll 3{H}|\dot\phi|$, on equations (\ref{FM2}) and (\ref{KG2}), which thus finally reduce to,
\begin{align}-54\beta H^4\left({{k^2-1}\over k^2}\right)^2-3\alpha H^2\left({{k^2-1}\over k^2}\right)+V(\phi) =0,~~~~~~3{H}\dot\phi \simeq - V'.\label{com2}\end{align}
Solving for $H^2$ in view of \eqref{com2}, we readily obtain,

\begin {align} H^2 = \frac{k^2\left[-\alpha\pm \sqrt{\alpha^2+24\beta V(\phi)}\right]}{36\beta (k^2-1)}.\label{soln2}\end{align}
Further, combining equations (\ref{com2}), one can  show that the `potential slow roll parameter' $\epsilon$ equals the `Hubble slow roll parameter' ($\epsilon_1$) under the condition,

\begin {align}\label{SR2}\begin{split} \epsilon \equiv - {\dot {H}\over {H}^2}=\frac{72\beta^2(k^2-1)V'^2}{k^2\left(\sqrt{\alpha^2+24\beta V(\phi)}\right)\left[-\alpha\pm \sqrt{\alpha^2+24\beta V(\phi)}\right]^2};\hspace{0.3 in}\eta = 2 \alpha \left({V''(\phi)\over V(\phi)}\right).\end{split}\end {align}
Further, since $\frac{H}{\dot\phi}=-\frac{k^2\left[-\alpha\pm\sqrt{\alpha^2+24\beta V(\phi)}\right]}{12\beta (k^2-1)V'}$, one can compute the number of e-folds as,

\begin{align}\label{Nphi2} {N}(\phi)\simeq \int_{t_i}^{t_f}{H}dt=\int_{\phi_i}^{\phi_f}{{H}\over {\dot\phi}}d\phi= \int_{\phi_f}^{\phi_i}\frac{k^2\left[-\alpha\pm\sqrt{\alpha^2+24\beta V(\phi)}\right]}{12\beta (k^2-1)V'}d\phi.\end{align}
Let us now make the same choice of the potential as in the case-1, viz., $V(\phi)= V_0-{V_1\over \phi}$, so that it remains almost flat as $\phi$ is large, and slow roll is admissible. Therefore, following two sets for the expressions of $\epsilon, ~\eta$ \eqref{SR2} and $N$ \eqref{Nphi2} are found,
\begin {align}\label{para2}\begin {split} &\epsilon =  \frac{72\beta^2(k^2-1){V_1}^2}{k^2\phi^4\left(\sqrt{\alpha^2+24\beta V_1\left(\frac{V_0}{V_1}-\frac{1}{\phi}\right)}\right)\left[-\alpha + \sqrt{\alpha^2+24\beta V_1\left(\frac{V_0}{V_1}-\frac{1}{\phi}\right)}\right]^2};\hspace{0.3 in}\eta =-\frac{2M_P^2}{\phi^2(\frac{V_0}{V_1}\phi-1)},\\& N=\int_{\phi_f}^{\phi_i}\frac{k^2\phi^2\left[-\alpha +\sqrt{\alpha^2+24\beta V_1\left(\frac{V_0}{V_1}-\frac{1}{\phi}\right)}\right]}{12\beta (k^2-1)V_1}d\phi.\end{split}\end{align}
\begin {align}\label{para21}\begin {split} &\epsilon =  \frac{72\beta^2(k^2-1){V_1}^2}{k^2\phi^4\left(\sqrt{\alpha^2+24\beta V_1\left(\frac{V_0}{V_1}-\frac{1}{\phi}\right)}\right)\left[-\alpha - \sqrt{\alpha^2+24\beta V_1\left(\frac{V_0}{V_1}-\frac{1}{\phi}\right)}\right]^2};\hspace{0.3 in}\eta =-\frac{2M_P^2}{\phi^2(\frac{V_0}{V_1}\phi-1)},\\& N=\int_{\phi_f}^{\phi_i}\frac{k^2\phi^2\left[-\alpha-\sqrt{\alpha^2+24\beta V_1\left(\frac{V_0}{V_1}-\frac{1}{\phi}\right)}\right]}{12\beta (k^2-1)V_1}d\phi.\end{split}\end{align}
 corresponding to a plus \eqref{para2} and minus \eqref{para21} sign appearing before the square root. In Table-2, we present a data set for the set of expressions \eqref{para2}, while Table-3 is presented for \eqref{para21}. In both the tables, we have varied ${V_0\over V_1}$ between $4.4 M_P^{-1} \leq\phi_i\leq 5.6 M_P^{-1}$, so that $r$ and $n_s$ lie within the experimental limit. Further, the number of e-folds for Table-2 and Table-3 are found to vary with in the range $48 \leq N \leq 56$ and $53 \leq N \leq 61$ respectively, which are sufficient to solve the horizon and flatness problems. Clearly, the inflationary parameters not only admit current observational constraints, the scalar to tensor ratio is able to sustain further constraints, which might appear from future analysis. The variation of $n_s$ with $r$ for the table 2 and table 3 are shown in figure 2 and figure 3 respectively.\\
\begin{figure}
\begin{minipage}[h]{0.47\textwidth}
      \centering
      \begin{tabular}{|c|c|c|c|c|}
     \hline\hline
      $\phi_f$ in $M_P$ &${V_0\over V_1}$ in $M_P^{-1}$ & $n_s$ & $r$ & $ {N}$\\
      \hline
        0.37422 &5.6 &  0.97128 & 0.00156 & 56\\
        0.38181 &5.4 &  0.97014 & 0.00166 & 54\\
        0.38990 &5.2 &  0.96889 & 0.00176 & 53\\
        0.39856 &5.0 &  0.96755 & 0.00188 & 52\\
        0.40787 &4.8 &  0.96607 & 0.00201 & 51\\
        0.41788 &4.6 &  0.96447 & 0.00216 & 50\\
        0.42872 &4.4 &  0.96270 & 0.00233 & 48\\
    \hline\hline
    \end{tabular}
     \captionof{table}{Data set \eqref{para2} for the inflationary parameters with $\phi_i=3.0 M_P$, $\beta V_1=1.0 M_P^5$,~$\alpha=0.5 M_P^2$ and $k=1.08$ under the variation of ${V_0\over V_1}$.}
      \label{tab:Table2}
   \end{minipage}%
  \hfill%
 \begin{minipage}[h]{0.47\textwidth}
\includegraphics[ width=1.10\textwidth] {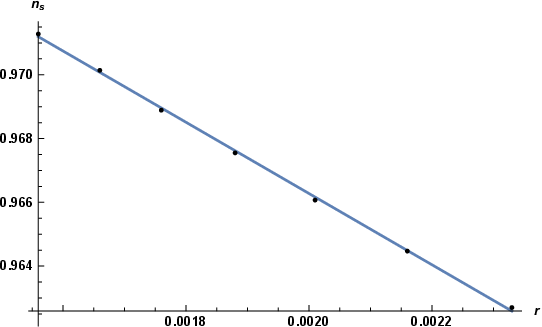}
 \caption{Variation of $n_s$ with $r$ as per data set of \eqref{para2}.}
      \label{fig:2}
   \end{minipage}
\end{figure}

\begin{figure}
\begin{minipage}[h]{0.47\textwidth}
   \centering
   \begin{tabular}{|c|c|c|c|c|}
     \hline\hline
      $\phi_f$ in $M_P$ &${V_0\over V_1}$ in $M_P^{-1}$ & $n_s$ & $r$ & $ {|N|}$\\
      \hline
        0.35797 &5.6 &  0.97138 & 0.00130 & 61\\
        0.36494 &5.4 &  0.97024 & 0.00138 & 60\\
        0.37237 &5.2 &  0.96901 & 0.00146 & 58\\
        0.38031 &5.0 &  0.96767 & 0.00156 & 57\\
        0.38883 &4.8 &  0.96621 & 0.00166 & 56\\
        0.39780 &4.6 &  0.96461 & 0.00177 & 55\\
        0.40790 &4.4 &  0.96286 & 0.00190 & 53\\
    \hline\hline
    \end{tabular}
     \captionof{table}{Data set \eqref{para21} for the inflationary parameters with $\phi_i=3.0 M_P$, $\beta V_1=1.0 M_P^5$,~$\alpha=0.5 M_P^2$ and $k=1.08$ under the variation of ${V_0\over V_1}$.}
      \label{tab:Table3}
     \end{minipage}%
     \hfill%
  \begin{minipage}[h]{0.47\textwidth}
\includegraphics[ width=1.10\textwidth] {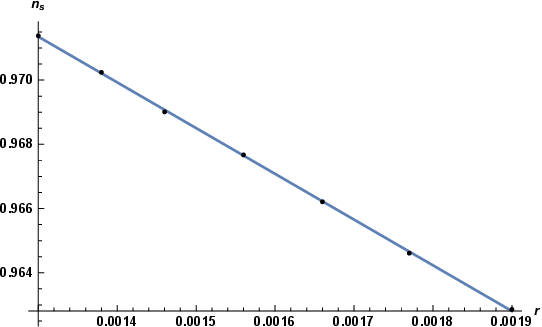}
 \caption{Variation of $n_s$ with $r$ as per the data set of \eqref{para21}.}
      \label{fig:3}
   \end{minipage}
\end {figure}

Let us now compute the energy scale of inflation in view of the relation \eqref{soln2}, considering the last but one data, ($N = 50$, for which $\phi_i = 3.0M_P,~ \alpha=0.5,~k=1.08, ~{{V_0\over V_1}=4.6 M_P^{-1}},~\beta V_1= 1.0M_P^5$), appearing in Table 2. Correspondingly, we compute,

\begin{align}\label{ES2}{H_*}^2 = \frac{1.1664\left[-\alpha\pm\sqrt{\alpha^2+24\beta V_1\left({V_0\over V_1}-{1\over \phi_i}\right)}\right]}{36\beta(1.1664-1)}=\frac {1.1664(-0.5\pm10.13)}{-5.990\beta} = \frac{1.875}{\beta},\end{align}
where positive value of ${H_*}^2$ is taken. Note that $\beta$ still remains arbitrary. Now, the energy scale of inflation in a single scalar field model corresponding to GTR \cite{Wands} is given by the following expression,

\begin{align}{H_*} = 8\times 10^{13}\sqrt{\frac{r}{0.2}}GeV = 8.314\times 10^{12} GeV \approx 3.34\times 10^{-6}M_P.\label{60},\end{align}
whose numerical value is computed taking into account the value of the tensor-to-scalar ratio $r = 0.00216$ from the last but one data set of Table 2. Thus, in order to match the scale of inflation \eqref{ES2} with the single field scale of inflation \eqref{60}, $\beta$ has to be, $\beta \approx 1.681 \times 10^{11} $. We have therefore been able to fix the parameter $\beta$, from physical ground (sub-Planckian inflation), and hence the values of $V_1$ and $V_0$ may be found as well, which are,

\begin{align} \label{V2}V_1= \frac{1.0}{1.68\times 10^{11}}\approx 5.95\times 10^{-12}M_P^5, \hspace{0.5 in} V_0 = 4.6 V_1\approx 2.74\times 10^{-11}M_P^4.\end{align}

Finally, to handle the issue of gracefully exit from inflation, we recall equation \eqref{FM2}, which in view of the above form of the potential, $V(\phi)=V_0-{V_1\over \phi}$, is expressed as,

\begin{align}-\frac{54\beta H^4(k^2-1)^2}{V_1k^4} - \frac{3\alpha H^2(k^2-1)}{V_1k^2}+\left[{\dot\phi^2\over 2V_1}+\left({V_0\over V_1}-{1\over {\phi}}\right)\right]=0.\end{align}
During inflation, $H^2$ and $V_1$ are of the same order of magnitude, while the Hubble parameter varies slowly. But, at the end of inflation, the Hubble rate usually decreases sharply, and $\beta H^4$ falls much below $V_1$. Hence, one can neglect both the terms ${\beta H^4\over V_1}$ and ${\alpha H^2\over V_1}$  without any loss of generality. Thus, one can write

\begin{align}{\dot{\phi}}^2=-2\left[V_0-{V_1\over \phi}\right].\label{osc2}\end{align}
Taking into account, $\phi_f=0.41788M_P$, $V_0= 2.74\times 10^{-11}M_P^4$ and $V_1= 5.95\times 10^{-12}M_P^5$ from Table-2, it is possible to show that the above equation exhibits oscillatory behavior

\begin{align}\phi=\exp({i\omega t}),\end{align}
provided, $\omega \approx 1.23\times10^{-5}M_P$. It may be mentioned that such astounding match of the inflationary parameters with the observational ones, was also found for $f(\mathbb{T}) = \alpha \mathbb{T} + \beta {\mathbb{T}}^2$ teleparallel gravity theory in isotropic and homogeneous model \cite{ftwe}. As mentioned in the introduction, $f(\mathbb{T})$ and $f(Q)$ in coincidence gauge (adopted here) are indistinguishable in homogeneous and isotropic model, so the result holds for $f(Q)$ gravity, as well.

\section{Radiation era} \label{sec4}

No doubt, the $f(Q)$ theory admits the observational constraints of the latest released inflationary data with extremely good precession. It is therefore required to scrutinize its validity in the subsequent epoch. After graceful exit from inflation, the universe re-heats and reaches the temperature ($\sim 10^{32}~K$) of hot big-bang, whence the universe enters the radiation dominated era. In this era, The seeds of perturbation which enter the horizon at the end of inflation, evolve to form structure. Further, CMB is also formed in this epoch. These are possible only if the universe evolves with decelerated expansion. So let us seek a solution in the form,

\be \label{17}a(t) = e^\xi = a_0 t^n, ~\mathrm{i.e.}~ \dot \xi = {n\over t}.\ee
Next, let us inspect under what condition the Klein-Gordon equation \eqref{7} is validated, which currently takes the following form, for the same choice of the potential $V(\phi) = V_0 - {V_1\over \phi}$, as

\be\label{18} \ddot \phi + 3{n\over t} \dot \phi + {V_1\over \phi^2} = 0.\ee
The above equation is satisfied for $l = {2\over 3}$, if we seek a solution in the form $\phi = \phi_0 t^{l}$. This means $\phi\propto t^{2\over 3}$, and $\phi$ increases with time. Nothing to worry about it, since $V_1 \sim 10^{-11}~M_P^5$ \eqref{V1} , and $V_1 \sim 10^{-12}~M_P^5$ \eqref{V2}, and therefore, soon the last term $V_1\over \phi^2$ becomes insignificantly small and may be neglected, leaving only $\ddot \phi + 3{n\over t} \dot \phi = 0$, behind. This admits a solution in the form

\be\label{phi1}\phi = \phi_0 t^{1-3n},\ee
and so for $3n < 1$, the remnant of $\phi$ is red-shifted away and there might not remain any trace of a scalar field in the low energy limit. In this manner the cosmic puzzle may be solved through $f(Q)$ gravity, avoiding dark energy issue.\\

\subsection{$f(Q) \propto Q^m$:}

We have already noticed that $f(Q) = -\beta Q^2$ also admits observational constraints on the inflationary parameters with excellence. So, in this subsection, we make our way to test its behaviour in the radiation dominated era, starting from an arbitrary power $m$. So, let us take up equation \eqref{4}, viz., $f_Q\dot \chi = c_1 e^{-3\xi}$, and consider the earlier assumption $\dot\xi = k\dot\chi$ \eqref{5} to obtain:

\be\label{19} f_Q\dot \xi = kc_1 e^{-3\xi},\ee
which in view of \eqref{17}, implies

\be \label{20} f_Q = {kc_1\over na_0^3}t^{1-3n}, ~~\mathrm{i.e.},~f_Q \propto {1\over t^{3n-1}}.\ee
Further considering, $f(Q) = f_0 Q^m$, we find

\be \label{20a} f_Q = mf_0 Q^{m-1} = mf_0 [-6(k^2-1)]^{m-1} \dot \chi^{2(m-1)} = mf_0 \left[-{6(k^2 - 1)\over k^2}\right]^{m-1} \dot \xi^{2(m-1)}; ~~\mathrm{i.e.},~ f_Q \propto {1\over t^{2(m-1)}},\ee
using solution \eqref{17}. Thus, equations \eqref{20} and \eqref{20a} simultaneously hold provided, $3n-1 = 2m-2$ or $n = {1\over 3}(2m-1)$. Now for $m = 1,~n = {1\over 3}$, and $f(Q) = f_0 Q$, which is simply `General Theory of Relativity'. The effective scale factor therefore evolves as $a(t) = a_0 t^{1\over 3}$, while the scalar field seizes to evolve in view of \eqref{phi1}. Note that in reference to equation \eqref{rhop2}, the potential vanishes. This is the solution obtained earlier with viscous fluid in the B-I space-time, in the absence of a scalar field \cite{B3}. This gives slower expansion rate than the standard Friedmann model ($a \propto \sqrt t$). On the contrary, if we seek $a \propto \sqrt t$, as in the Friedmann model, then $m = {5\over 4}$, i.e. $f = f_0 Q^{5\over 4}$, which is also not much promising. Particularly for $m = 2$, which we have considered to exhibit outstanding fit with observed inflationary parameters, leads to coasting solution $a \propto t$ and for $m > 2$ accelerating expansion is revealed. Therefore, despite admitting observational constraints on inflationary parameters with excellence, $f(Q) \propto Q^2$ is not a good choice to explain radiation dominated era.\\

\subsection{$f(Q) = \alpha Q - \beta Q^2$:}

From the above analysis, it is clear that the assumption $\sigma^2 \propto \theta^2$ does not allow a combination of two terms ($f(Q) = \alpha Q - \beta Q^2$) as considered in the present subsection. We therefore withdraw this assumption as a result of which $\xi(t)$ and $\chi(t)$ are no longer related. Hence, we are required to solve the set of equations \eqref{6}. Again we start with our master equation \eqref{4} as before, viz, $f_Q\dot \chi = c_1 e^{-3\xi}$, and use solution \eqref{17} to find,

\be\label{21} \alpha\dot\chi + 12 \beta \left({n^2\over t^2} - \dot\chi^2\right)\dot\chi - {c_1\over a_0^3 t^{3n}} = 0.\ee
Equation \eqref{21} is satisfied under the condition:

\be\label{22} \dot \chi = {\chi_0\over t};~ \chi_0 =  n = {1\over 3};~\alpha = {c_1\over a_0^3\chi_0} = {3c_1\over a_0^3};~a= a_0 t^{1\over 3}.\ee
Note that in the process we arrive at $\dot\chi^2 = \dot\xi^2$, leading to $Q = 0$. Apparently, it does not create any problem, since a viable radiation era ($a \propto \sqrt t$) of standard cosmology requires the Ricci scalar $R = 0$. Nonetheless, we have already noticed in subsection (3.1), that the energy conditions are satisfied provided $|\alpha Q| > 2\rho_\phi$, and $|\alpha Q| > \rho_\phi + p_\phi$. Therefore, $Q = 0$ clearly violates energy conditions. Further, for $n = {1\over 3}$, the scalar field \eqref{phi1} seizes to evolve, and so it would appear as dark energy in the matter dominated era. In this process we get tangled up, since all the development in regard of symmetric telleparallel theory to resolve cosmic puzzle without dark energy ended in the requirement of dark energy, yet again. Further, note that the chosen form of $F(Q)$ is a minimal extension of GTR, while other forms make things more complicated. Next, we observe that even after withdrawing the assumption $\sigma^2\propto\theta^2$, a decelerated expansion requires $Q=0$. It is noteworthy that in GTR, the Ricci scalar $R=0$ does not mean vanishing of curvature, since curvature is determined by the Kretchmann scalar $R_{\alpha\beta\gamma\delta}R^{\alpha\beta\gamma\delta}$. Nonetheless, in non-metricity theory under consideration, no such scalar is known to exist, and so $Q=0$ implies collapse of the nonmetricity teleparallel theory itself. Therefore, we conclude that radiation dominated era does not yield a decelerated expansion, for the usual connection that has been considered here, if the scalar field is completely used up in the process of creating particles at the end of inflation. However, we find it extremely difficult to find solution in the presence of the scalar field. It is also noteworthy that other connections exist for Bianchi-I space-time, yielding a different form of $Q$ altogether, and emerging other results. Therefore, it is too early to conclude that $f(Q)$ theory of gravity fails to produce a viable cosmological evolution in the radiation era.

\section{\textbf{Concluding remarks}}\label{sec5}

Resolving the cosmic puzzle without dark energy, requires to modify GTR. Several such modified theories of gravity have been developed in the last two decades. Recent hearsay being the `teleparallel gravity theory'. The primary advantage of `teleparallel gravity theory' is that the field equations are second order unlike $f(R)$ theory of gravity. This avoids Ostrogradski instability and associated ghost degrees of freedom. However, Lovelock gravity also has the same characteristics, but, it is applicable either in higher dimensions ($D > 4$), or requires to associate a dilatonic coupled scalar field in $4$ dimension. In teleparallel gravity, no such field is required to explain late-time cosmic puzzle. However, to drive inflation in the very early universe, a dilatonic or Higgs scalar field may be associated, which must have been redshifted to a large extent today. This was our motivation to associate a scalar field. Two types of teleparallism have been investigated in recent years, the metric $f(\mathbb{T})$ and the symmetric $f(Q)$ teleparallel gravity theories. In the present article we have analysed the latter theories to demonstrate the evolution of the universe ranging from very early to late accelerating stage. The primary reason has been discussed largely in the introduction.\\

We have made the following observations after a thorough investigation of the comparatively new $f(Q)$ gravity:
\begin{itemize}
\item In the present article we have studied $f(Q)$ theory of gravity in the anisotropic yet homogeneous axially symmetric Bianchi-I model, under the coincident gauge coordinates. Although it administers latest released observational constraints on inflationary parameters with outstanding excellence, unfortunately, it miserably fails to envisage a decelerated expansion in the radiation dominated era, which follows inflationary regime and is responsible to produce CMBR.
\item It might appear that the claim (it miserably fails to envisage a decelerated expansion in the radiation dominated era) is not general, since, on one hand we have worked with a particular form of $f(Q)$, being a minimal extension of GTR, and on the other, several physically viable assumptions have been made to find exact solutions. Nonetheless, in section (5.2) we have observed that even after withdrawing the assumption $\sigma^2 \propto \theta^2$, for $f(Q) = \alpha Q- \beta Q^2$, a decelerated expansion requires $Q = 0$. It is noteworthy that in GTR, the Ricci scalar $R = 0$ does not mean vanishing of curvature. Since $R$ being the trace of Ricci curvature tensor, measures the sum of all sectional curvatures of planes spanned by distinct pairs of elements in a given orthonormal basis. Nonetheless, in nonmetricity theory under consideration, the scalar $Q$ is not the trace of nonmetricity tensor $Q_{\alpha\beta\gamma}$, and so $Q = 0$ appears to imply collapse of the current $f(Q)$ theory itself. Even if not, atleast we observe the violation of energy conditions of thermodynamic fluids. It is important to mention that, $Q = 0$ is the primary requirement for decelerated expansion in the radiation dominated era, and hence, it does not depend on the choice of $f(Q)$. Therefore, we conclude that radiation dominated era does not yield a decelerated expansion. Further, the solution so obtained with $Q = 0$, freezes the scalar field to evolve any further, and so it remains intact during matter dominated era. This scalar acts as a quintessence field. Thus the main aim of teleparallel theory, which is to explain late-time cosmic evolution with accelerated expansion, without dark energy, fails.
\end{itemize}
However, it is too early to discard the `symmetric teleparallel theory', since the coincident gauge connections that we have worked with, and most other literature also adapted to, is not the unique connection to formulate the $f(Q)$ theory \cite{flrwconn}. In future works we hope to derive all the connections in the Bianchi type-I spacetime which are compatible with symmetric teleparallelism, and associate shear and bulk viscosities, to further investigate the effect in the early (inflationary epoch) and mid-stage evolution (radiation dominated epoch) of the universe.

\section{\textbf{Appendix}}\label{appI}
Let us consider a quadratic model $f(Q)=\lambda Q^2$. In radiation era ($p=\frac13 \rho$), from the energy equation and one of the pressure equations (\ref{1}), we have the relation
\begin{equation}\label{a2}
    Q(\dot{\chi}+\dot{\xi})=c_3e^{-3\xi}\,.
\end{equation}
Whereby, using equation (\ref{4}) in (\ref{a2}), we have
\begin{equation}
    \xi=k\chi+k_1,
\end{equation}
where $k$ and $k_1$ are some constants. This linear relation between $\chi$ and $\xi$ that we obtained is remarkable, as it gives a much accepted physical relation, the expansion scalar $\theta=3\dot{\xi}$ is proportional to the shear$\sigma$,where, $\sigma^2=3\dot{\chi}^2$.

\end{document}